\documentclass[aps,
 amsmath,amssymb,
 prl,
 reprint,
 superscriptaddress
 ]{revtex4-2}

\usepackage{graphicx}
\graphicspath{{img/}}
\usepackage{dcolumn}
\usepackage{bm}
\usepackage{hyperref}
\usepackage{cleveref}

\usepackage{multirow}
\usepackage{amsmath,amssymb,amsfonts}
\usepackage{amsthm}
\usepackage{mathrsfs}
\usepackage[title]{appendix}
\usepackage{xcolor}
\usepackage{booktabs}
\usepackage[separate-uncertainty=true,multi-part-units=single]{siunitx}

\usepackage[utf8]{inputenc}
\usepackage[T1]{fontenc}

\begin{document}

\title{Using tunable coherence for reaching micrometer coherence lengths\\ and suppressing stray light in a power-recycled Michelson interferometer}

\author{Daniel Voigt}
\email[]{daniel.voigt@uni-hamburg.de}

\author{Oliver Gerberding}
\email[]{oliver.gerberding@uni-hamburg.de}
\affiliation{Institute for Experimental Physics, 
University of Hamburg, 
Luruper Chaussee 149,
22761 Hamburg,
Germany}

\date{\today}

\begin{abstract}
By reentering into laser interferometers, scattered or stray light introduces non-linear noise. This is a major limitation of precision interferometers as preventing such parasitic light is nearly impossible. 
Thus, substantial effort is put into mitigating the reentering of these fields in various ways. Ground-based laser interferometric gravitational wave detectors employ such mitigation techniques to reduce otherwise restrictive stray light noise. 
However, they are now reaching sensitivities where conventional mitigation techniques reach limitations. Further improvements planed for future observatories are placing even more demanding constraints on tolerable stray light power. 
We previously presented tunable coherence as a possible technique to ease these constraints and suppress unwanted coherent interference. 
For these promising demonstrations, the remaining coherence length and achievable suppression in length-constrained layouts was limited, among other things, by the used pseudo-random-noise phase modulation frequency. 
In this work, we demonstrate stray light suppression and cavity performance at modulation frequencies up to \SI{10}{\giga\hertz}. 
This reduces the remaining coherence to a few centimeter in an interferometer, and even to the scale of the laser wavelength in a cavity. 
We further present a first demonstration of tunable coherence in a power-recycled Michelson interferometer, successfully suppressing stray light in a more complex topology.
\end{abstract}

\maketitle

\section{Introduction}\label{sec:introduction} 

Scattered or stray light is a significant noise source in high precision laser interferometers~\cite{schilling1981,schnupp1985,accadia2010,ringlaser_Korth2015,martynov2016,buikema2020,capote2025,ringlaser_igel2021,ringlaser_schreiber2023}. 
Its creation due to light of the main laser being scattered out of the intended path is impossible to fully avoid. 
If this light reenters the interferometer, it introduces noise by interfering with nominal light inside the interferometer. This can also lead to opto-mechanical disturbances by inducing radiation pressure fluctuations. 
Even though the unintended path can be long, the light can still interfere due to the
high coherence that is normally desired for high precision experiments. 
Any reflection from a dynamic object then introduces parasitic phase information which couples non-linearly into the readout~\cite{schilling1981,vinet1996}. Especially if the introduced dynamics exceed the wavelength of the laser, this can lead to frequency up-conversion and broadband noise~\cite{vinet1996,accadia2010}. 

A good example for this are ground-based gravitational wave detectors. 
These kilometer-scale interferometers are designed to measure relative length deviations of less than $10^{-21}$ in their most sensitive bandwidth around some hundreds of hertz. 
The current network of observatories~\cite{LVK_Network}, consisting of the Advanced LIGO~\cite{LIGO}, Advanced VIRGO~\cite{VIRGO}, KAGRA~\cite{KAGRA} and GEO600~\cite{GEO600} detectors, steadily increases the rate of detected events through ongoing improvements of the interferometers. 
The next generation of observatories like the Einstein Telescope~\cite{ET,ETbluebook2025} and Cosmic Explorer~\cite{CosmicExplorer,CosmicExplorer2} are designed to further increase the sensitivity. Here, especially the former aims to also extend the peak sensitivity down to 3\,Hz where current detectors are limited by several noise sources. 
One of the leading contributions to this will be stray light~\cite{martynov2016,buikema2020,capote2025}, with the full extend of it currently only partially understood~\cite{capote2025}. 

Stray light mitigation has always been an integral part of detector development~\cite{schilling1981,vinet1997,ottaway2012,chua2014,soni2021,nguyen2021,Soni2024}. 
Most commonly used for this are baffles to block and in some cases measure~\cite{ballester2022,andrs-Carcasona2025} the stray light. Others include control adaptations~\cite{nguyen2021}, dampening or changing of scatterer motions~\cite{luck2008,soni2021,Soni2024} or the use of additional readout information~\cite{ast2016,Was2021,Acernese2022}.
However, even at the current sensitivities, these are not enough as e.g. baffles become the source of stray light themselves~\cite{soni2021,capote2025}. 
In previous works~\cite{voigt2023,voigt2025,eggers2025}, we demonstrated the concept \textit{tunable coherence} as a possible solution for stray light mitigation. 
The approach, using pseudo-random-noise (PRN) phase modulation, had been considered already during early detector development but was not yet technically feasible then~\cite{schnupp1985,dewey1986}.

The concept tunable coherence aims to directly address the problem of stray light by inhibiting its ability to interfere. 
In contrast to other techniques, this approach can work without detailed knowledge of scatter sources as it reduces the coherence of any stray light and thus prevents its interference with the interferometer. 
For this, binary PRN sequences are phase modulated onto the laser at frequency $f_{\text{PRN}}$, introducing phase flips of $\pi$. 
This restricts interference between modulated light beams by randomizing their phase relation. 
The ability of PRN modulated light to interfere is thus characterized by two new lengths. One is the length of a single chip in the binary sequence that determines its coherence length $d_{\text{chip}}\!=\!c/f_{\text{PRN}}$, where $c$ is the speed of light. The other is the optical path length of the whole sequence, after which it repeats. This is given by $d_{\text{coh}}\!=\!n_{\text{chips}}d_{\text{chip}}$, where $n_{\text{chips}}$ is the number of chips in the sequence. 
In previous works, the effects of tunable coherence in different interferometer topologies~\cite{eggers2025,voigt2025} and an optical resonator~\cite{voigt2025} were simulated~\cite{voigt2023} and experimentally demonstrated at \SI{1}{\giga\hertz} modulation frequency. 
We now present results obtained with PRN frequencies up to \SI{10}{\giga\hertz}, reducing the remaining coherence to below \SI{5}{\centi\meter} and even to wavelength scales in optical cavities. Further, we also demonstrate tunable coherence in a power-recycled Michelson interferometer (PRMI) for the first time, moving towards the more complex optical topologies of real gravitational wave detectors.

\section{Results}\label{sec:results} 
\begin{figure}[!t] 
    \centering
    \includegraphics[width=0.48\textwidth]{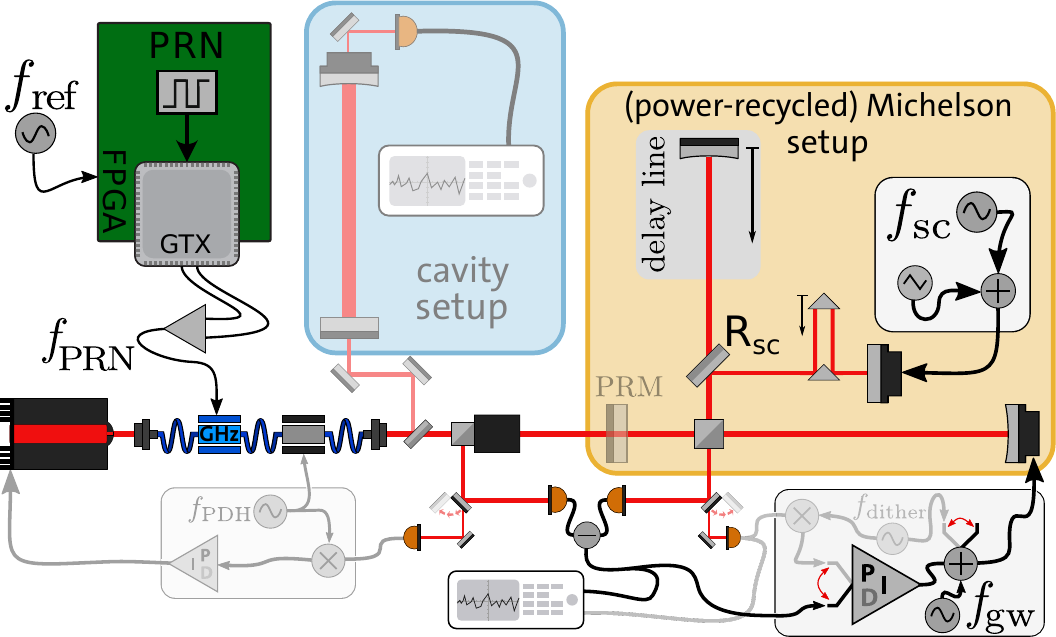}
    \caption{Sketch depicting the individual parts of the experimental setup. The laser preparation with two EOMs, one for modulating the PRN sequence onto the light, and one for rf-sidebands used for the PDH-technique. The Michelson interferometer setup is depicted in the orange box with the possible addition of power-recycling. Without recycling, the interferometer was operated and read out at mid-fringe. Power-recycling case was realized by operating the interferometer at dark-fringe using a dither-lock and adding a PRM before the interferometer. The resulting power-recycling cavity was operated by locking the laser frequency to its resonance using the PDH-technique. The stand-alone cavity depicted in the blue box was operated separately. Its microscopic length was adjusted to keep it on resonance with the laser frequency using a PZT mirror and the PDH-technique.}
    \label{fig:setup_sketch}
\end{figure}

In previous demonstrations the required precision for matching optical delays between intended interference needed no dedicated equipment~\cite{voigt2025}. 
By increasing the PRN frequency up to \SI{10}{\giga\hertz}, we now demonstrate an order of magnitude reduction in remaining coherence length. 
This reduces the amount of stray light that can interfere within the setup, however, it also tightens the requirements for length matching significantly. 
Our setup therefore not only contained piezo-actuated (PZT) elements for microscopic but also for macroscopic length tuning. All parts of the setup are depicted in Fig.~\ref{fig:setup_sketch} and described in more detail in the End Matter. 

\subsection{Michelson interferometer}\label{sec:michelson}
The first experiment used a Michelson interferometer depicted in the orange box of Fig.~\ref{fig:setup_sketch}. One of the arms contained a PZT mirror for adjusting the phase tuning between the arms, the other contained a PZT delay line to adjust the macroscopic length offset between them. Both arms were matched in length and the interferometer locked to a mid-fringe operating point. Stray light coupling was emulated by picking of light in one of the arms with a low-reflectivity beam splitter ($R_{\text{sc}}\!\approx\!\SI{20}{\percent}$). This light traveled an additional, variable delay, before being back-reflected by another PZT mirror to modulate a signal at $f_{\text{sc}}$ onto it. It then reentered the interferometer at the same beam splitter. 

\begin{figure*}
    \centering
    \includegraphics[width=0.95\textwidth]{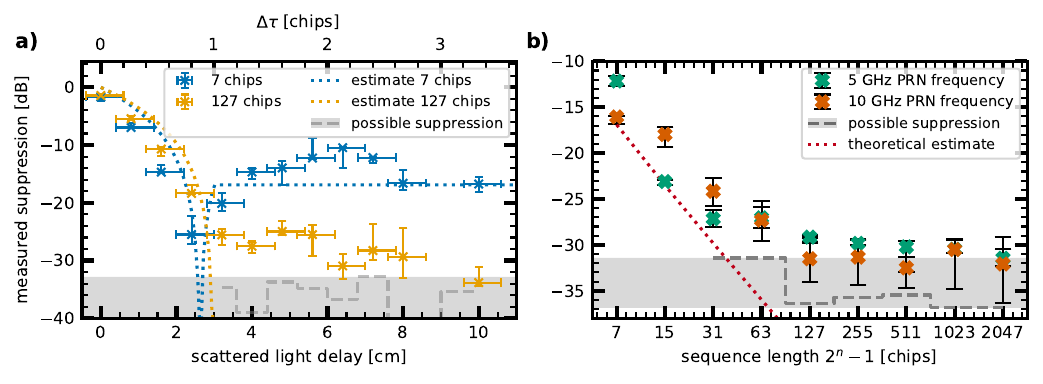}
    \caption{Measured stray light suppression in a Michelson interferometer using tunable coherence. Figure a) shows the dependence on the delay of the scattered light, measured for a PRN sequence length of \num{7} and \num{127}~chips at $f_{\text{PRN}}\!=\!\SI{10}{\giga\hertz}$. Both sequences reached either their expected suppression or the experimental limit at the longest delay of \SI{10}{\centi\meter}. For shorter delays they both showed some fluctuations and slower reduction of coherence for the longer sequence. Figure b) shows the dependence of measured suppression on sequence length for PRN frequencies of \num{5} and \SI{10}{\giga\hertz}. The shorter sequences showed some limitations, the longer sequences, starting with \num{127}~chips in length, reach the experimental limits but residual stray light often remains.}
    \label{fig:MichelsonResults}
\end{figure*}
We measured the achieved stray light suppression depending on two parameters: relative scattered light delay and length of the PRN sequence. Several sequences were stored on an FPGA, allowing for lengths between \num{7} and \num{2047} chips in steps of $2^n\!-\!1$. 
The stray light delay could be varied between \num{0} and \SI{10}{\centi\meter}. 
The results for both measurements are shown in Fig.~\ref{fig:MichelsonResults}, together with the expected residual coherence based on ideal, rectangular PRN modulation shapes with a modulation depth of $\pi$. This estimate is given by the auto-correlation of the sequence
\begin{equation}\label{eq:suppEstimate}
    R(t,\tau)
		= \left\{ \begin{array}{cl} 1 	&\text{if $\tau = 0$ (mod $t_{\text{seq}}$)} \\
			-\frac{1}{n_{\text{chips}}} &\text{if $\tau \neq 0$ (mod $t_{\text{seq}}$)} \end{array}\color{white}\right\}
\end{equation}
which for delays $\tau$ between $\pm t_{\text{chip}}\!=\!f_{\text{PRN}}^{-1}$ is described by
\begin{equation}
    R(|\tau|\!\leq\!t_{\text{chip}}) = 1 - \frac{|\tau|}{t_{\text{chip}}}\left(1+\frac{1}{n_{\text{chips}}} \right).
\end{equation}

In Fig.~\ref{fig:MichelsonResults}a we show the suppression measured with the \num{7} and \num{127}~chips long sequences depending on the stray light delay at $f_{\text{PRN}}\!=\!\SI{10}{\giga\hertz}$. Both sequences reached either their theoretical estimate or the maximal achievable suppression, determined by the strength of the emulated stray light above the noise floor of the given setup, at the longest delay of \SI{10}{\centi\meter}. For short delays, both sequences follow the theoretical estimate. However, the data showed a static offset of \SI{2}{\milli\meter} most likely coming from the calibration of the zero-delay point. To account for this, we extended the error bars towards one side by the offset to include this systematic error in the calibration. At a delay shortly below the length of one chip, the theoretical suppression estimate becomes infinite as the auto-correlation of the sequences cross zero. For the shorter sequence, we measured this stronger than usual suppression for a delay of \SI{2.4}{\centi\meter} at \SI{25.5}{\decibel}. Between a delay of \num{1} and \num{2.5}~chips, the suppression then deteriorates before fully agreeing with the estimate for the last two data points at a delay of \num{8} and \SI{10}{\centi\meter} or around \num{3}~chips. 
For the longer sequence, both the nominal expected maximum suppression, as well as the short span with increased suppression potential, were beyond our experimental limitations. Thus, instead the already in previous experiments observed~\cite{voigt2025} washed out approaching of the experimental limit with increasing delay was observed. 
At a delay of one chip, the performance started to deviate from expectations. Here it reached a suppression of around \SI{25}{\decibel} before flattening out slowly towards the experimental limit. Only for delays exceeding two chips, or about \SI{6}{\centi\meter}, it then reached this limit. 

The dependence of reached suppression on sequence length is shown in Fig.~\ref{fig:MichelsonResults}b for measurements at \num{5} and \SI{10}{\giga\hertz} PRN frequency. Here, the shorter sequences often fell slightly short of reaching full expected suppression. Only at lengths of \num{127}~chips and longer, the suppression reached the experimental limits, however, some stray light often remained. Notably, the \num{63}~chips long sequence, which was an outlier in previous works~\cite{voigt2025}, did suppress stray light more reliably in this reworked setup. The maximum suppression was achieved with the \num{511} and \num{2047}~chips long sequences at slightly above \SI{32}{\decibel}. In this range, the suppression reached the limitations and noise floor of our setup. 

\subsection{Power-recycled Michelson interferometer}\label{sec:PRMI}
As a next step towards more complex interferometer configurations, a so-called power-recycling mirror (PRM) was added into the input path before the main beam splitter of the interferometer. By operating the interferometer at darkfringe, meaning all light leaves the output towards the laser, it effectively forms a mirror. Together with the PRM, this results in a cavity used to increase the circulating power inside the interferometer. For this experiment, the reflectivity of the beam splitter for stray light emulation was reduced to about $R_{\text{sc}}\!\approx\!\SI{3.25}{\percent}$ to reduce losses inside the cavity. To compensate for the remaining losses, the reflectivity of the PRM was chosen to be \SI{90}{\percent}. From previous results~\cite{voigt2023,voigt2025}, it was expected that the precision for the length matching inside this cavity would be even stricter due to multiple roundtrips enhancing any mismatch~\cite{voigt2025}. 

\begin{figure}
    \centering
    \includegraphics[width=0.48\textwidth]{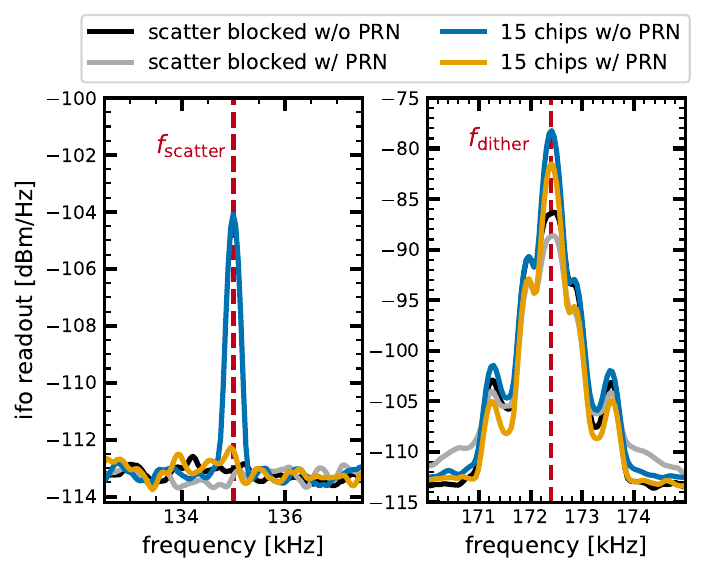}
    \caption{Power-spectral density recorded at the output of the power-recycled Michelson interferometer. On the left side the measured scattered light signal power and on the right side the measured dither signal power are shown. 
    Recordings with the scattered light blocked are plotted in black and gray, the recording without PRN modulation is plotted in blue and the recording with the modulation active in orange. The stray light signal power is reduced by about \SI{8.3}{\decibel}, and the SNR between stray light and dither signal improved by about \SI{5}{\decibel} when using the PRN modulation.}
    \label{fig:PRMIresults}
\end{figure}
With tunable coherence, two macroscopic lengths now had to be additionally controlled precisely to ensure normal operation of the interferometer: the relative length difference between the two arms and the roundtrip length of the cavity, or equivalently the PRN sequence repetition length. For the former, the PZT delay line was used. For the latter, the PRN frequency was tuned to adjust the repetition length as the cavity had a fixed length. The laser frequency was controlled to match the microscopic phase tuning of the cavity with the Pound-Drever-Hall (PDH) technique. 
The setup experienced instabilities and inconsistencies due to an unstable laboratory environment. Data was therefore recorded in quick succession to minimize negative influences on the measurement. 

Fig.~\ref{fig:PRMIresults} shows recorded spectra of two different measurements, each with and without PRN modulation. In the one shown in gray and black, the scattered light was blocked. A small reduction in the dither signal injected directly into the interferometer can be observed. 
The most likely cause for this is a small macroscopic mismatch of the two arms, enhanced by the cavity. However, a length mismatch between this cavity and the PRN sequence could also have caused it. 
The second measurement shown in blue and orange demonstrated about \SI{8.28}{\decibel} suppression of the injected stray light signal. However, also the dither signal is again suppressed slightly. Thus, a more reliable measure is the increase in signal-to-noise ratio (SNR) between the two injected signals. This SNR was calculated by comparing the peaks of the signal injected into the interferometer and the (intentional) scattered light noise. The ratio between these improved by \SI{5}{\decibel} between measurements with and without PRN modulation. 


\subsection{Cavity performance}\label{sec:cavity}
The last experimental setup was a stand-alone linear cavity as depicted in the blue box of Fig.~\ref{fig:setup_sketch}. This was used to further investigate the effects of tunable coherence on an optical resonator. Especially the higher PRN frequencies and increased integer ratios $\alpha\!=\!l_{\text{cav}}/d_{\text{coh}}$ between cavity roundtrip length $l_{\text{cav}}$ and PRN sequence length $d_{\text{coh}}$ were of interest in this context. The cavity was originally matched to a 15~chips long sequence at \SI{1}{\giga\hertz}~\cite{voigt2025}. We previously demonstrated that $\alpha$ needs to be an integer for full recovery of normal cavity performance~\cite{voigt2023,voigt2025}.  With the increased PRN frequencies we now realized different integer values for $\alpha$ between \num{5} and \num{22}, and the half-integer value of \num{2.5}, by tuning the exact PRN frequency for sequences 7, 15 and 32~chips long. 

\begin{figure*}
    \centering
    \includegraphics[width=0.95\textwidth]{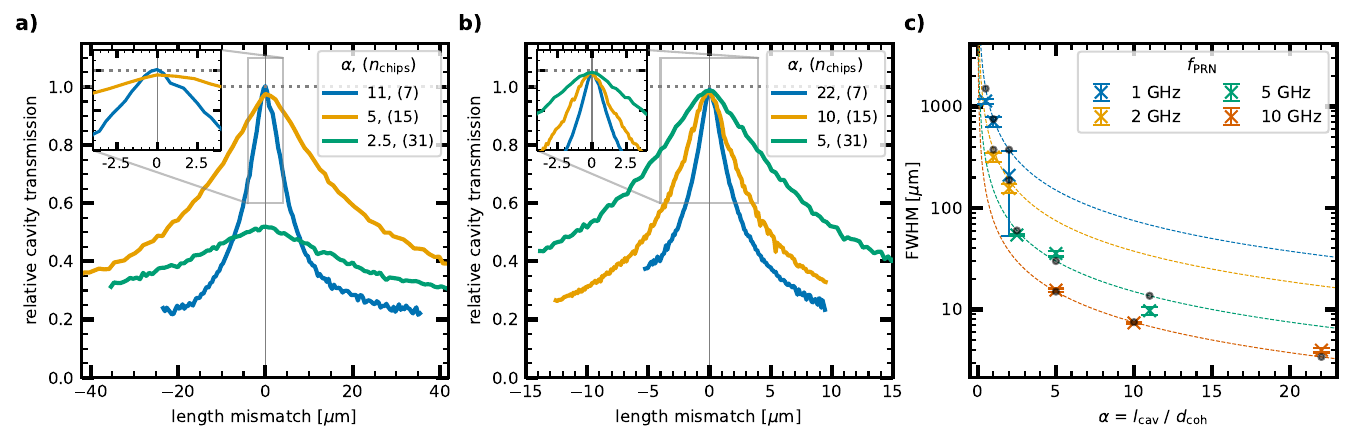}
    \caption{Measured macroscopic resonance of the cavity depending on the length matching between roundtrip length and PRN sequence length, shown for measurements using $f_{\text{PRN}}\!=\!\SI{5}{\giga\hertz}$ in figure a and $f_{\text{PRN}}\!=\!\SI{10}{\giga\hertz}$ in b. All combinations of integer multiples, given by $\alpha$, of sequence repetitions fitting into the cavity regain full resonance. With increasing $\alpha$, the resonance narrows. All measured FWHMs of the resonance are shown in figure c depending on $\alpha$ and PRN frequency.}
    \label{fig:CavityResults}
\end{figure*}
The results are shown in Fig.~\ref{fig:CavityResults}. Measurements were done at \num{5} and \SI{10}{\giga\hertz} PRN frequency, shown in Fig.~\ref{fig:CavityResults}a and b respectively. 
The figures show the recorded power in transmission of the cavity with PRN modulation active relative to the power recorded without the modulation. 

Fig.~\ref{fig:CavityResults}a shows the measurements at $f_{\text{PRN}}\!=\!\SI{5}{\giga\hertz}$, resulting in values for $\alpha$ of \num{2.5}, \num{5} and \num{11} for the longest to shortest sequence respectively. 
Both integer values resulted in full resonance, the half-integer in about half the transmitted power compared to the case without modulation. This is expected~\cite{voigt2023,voigt2025}.
Further, an increase in $\alpha$ led to a narrowing of the resonance. This is caused as the length mismatches of all sequence repetitions needed to reach the full cavity roundtrip length are added up.

The same can be observed in Fig.~\ref{fig:CavityResults}b which shows the measurements for the same sequence lengths but at $f_{\text{PRN}}\!=\!\SI{10}{\giga\hertz}$, resulting in $\alpha$ values of \num{5}, \num{10} and \num{22}. 
In both cases, the full-width-half-maximum (FWHM) of this resonance reduced to only tens of \si{\micro\meter} and even down to \SI{3.9(0.2)}{\micro\meter} for the highest $\alpha$ of \num{22}. 
All measured FWHMs are shown in Fig.~\ref{fig:CavityResults}c. 
We used the inverse relation of this FWHM observed for the PRN frequency $f_{\text{PRN}}$ and $\alpha$ and expected for the cavities finesse $\mathcal{F}$~\cite{voigt2023} to fit an empirical relation. 
With the so far measured values, such a fit resulted in 
\begin{equation}\label{eq:cavCoherence}
    \text{FWHM}_{\text{PRN}} = \frac{\pi}{2}\cdot\frac{c}{\alpha\cdot f_{\text{PRN}}\cdot \mathcal{F}}
\end{equation}
where the factor $\pi/2$ was found empirically through the fit. The speed of light $c$ is introduced through the 
proportionality of the FWHM to $d_{\text{chip}}\!=\!c/f_{\text{PRN}}$.

\section{Discussion}\label{sec:discussion} 
The measured suppression in the Michelson interferometer followed similar performance and limitation patterns as in previous work~\cite{voigt2025}. 
In the setup used for this work, the noise floor was higher, limiting the achievable suppression to about \SI{35}{\decibel}. This might be caused by the power imbalance resulting from the loss of light from the main interferometer at the optic coupling the stray light. Still, the suppression, which in most cases reached this experimental limit for long enough stray light delays, is very promising. With this work we reached remaining coherence lengths below \SI{5}{\centi\meter}, improving on previous results by about an order of magnitude. Thus, demonstrating once more that tunable coherence is experimentally feasible also on smaller scales. However, for fully taking advantage of high PRN frequencies to reduce the coherence length to its theoretical minimum, improvements in the modulation setup are needed. The limited bandwidth and missing precise control of the modulation depth degraded the performance. 
This is most relevant at high PRN frequencies, short distances, and when reaching high suppression as then already small deviations can spoil the needed precision. 

The first demonstration of tunable coherence in a PRMI is another promising step towards more complex layouts. However, further work is needed to fully show its potential in this context. The current implementation remains a proof of concept as general stability problems of the setup prevented full optimization for tunable coherence.
With a stable interferometer, more focus has to be put on properly matching all lengths and better investigating the influences of tunable coherence.

Finally, the very short coherence lengths achieved in optical resonators suggest that in such layouts also stray light with very small delays could be suppressed. 
The empirical relation given by equation~\ref{eq:cavCoherence} helps understanding this but also opens further questions: 
Based on our empirical model, for cavities significantly longer than the sequence repetition length or with high finesse, the effective coherence length could shrink down to, or even below, a wavelength. 
This could introduce significant challenges not only for the operation of the cavity but for its performance in general. 
High reflectivity mirrors relying on constructive interference from multiple layers of coatings could be penalized and a more complex model for these ultra-low coherence lengths has to be developed. On the other hand, it potentially also opens up new application areas. 
However, the naive remaining coherence, given by $d_{\text{chip}}$, will still be much larger. Thus, 
to which degree such effects actually occur is an important question for further studies.

\section{Conclusion}\label{sec:conclusion}
The results presented in this work demonstrate tunable coherence as a feasible technique to suppress stray light even in small scale interferometers. This opens even more applications than previously demonstrated~\cite{voigt2025,eggers2025} as the remaining coherence can be reduced to a few centimeter or even to micrometer lengths. With the demonstration of stray light suppression in a PRMI we now also demonstrated this ability of tunable coherence inside a resonator. 
However, to make full use of it, the bandwidth and control of the PRN modulation needs to be optimized. Then, table-top sized complex interferometers, using e.g. power-recycling, can be realistically combined with tunable coherence. Further, inside of optical resonators, the coherence length can even be reduced to the wavelength scale. 

For a possible implementation in gravitational wave detectors, the demonstrated increase in PRN frequency would allow for significant suppression using sequences short enough to fit e.g. the roundtrip length of the output mode cleaner cavities. With the long arm-cavities in these interferometers, the coherence length would then be expected to shrink even below a wavelength, further decreasing the number of stray light paths that could still interfere. However, whether such coherence lengths are actual realizable and what other implications and applications that could have, needs to be studied further. 
Another aspect to consider is the influence of tunable coherence on quantum noise in the detector. 
Here, the combination with squeezed light remains an open field for investigation. Currently, the extent of additionally, by the transitions between chips in the PRN modulation induced, phase noise that could limit the achievable squeezing factor was not yet quantifiable in our setup. We estimate, on the other hand, that the suppression of interference between stray light and nominal interferometer light also reduces radiation pressure noise. Here, we reason that the stray light contribution to this noise becomes broadband. Thus, the fluctuations are spread to the high PRN frequencies where the dampening of the pendulum stages is stronger.

Additionally, similar to digital interferometry~\cite{shaddock2007}, tunable coherence allows to distinguish between interferences from different beam sources. However, due to the concept working in the optical domain, higher spatial resolution can be realized using higher PRN frequencies~\cite{isleif2014} without high speed data acquisition as the modulation does not need to be resolved in the recorded data.

\section{Acknowledgments}
\begin{acknowledgments}
    The authors thank Mikhail Korobko for his valuable input during the LIGO P\&P process. 
    This research was funded by the Deutsche Forschungsgemeinschaft (DFG, German Research Foundation) under Germany's Excellence Strategy---EXC 2121 ``Quantum Universe''---390833306.
\end{acknowledgments}

\section{Appendix}
\appendix\label{sec:methods}
\subsection{PRN modulation generation}
The used laser was a non-planar ring oscillator (NPRO) \SI{1064}{\nano\meter} laser (Coherent: Mephisto 500) that was sent through a fiber coupled waveguide electro-optical modulator (EOM) (iXblue Photonics: NIR-MPX-LN-20) with a maximum bandwidth of \SI{20}{\giga\hertz}. So-called maximum-length-sequences (\textit{m}-sequences) generated by linear-feedback-shift-registers (LFSR) with varying lengths~\cite{alfke1996} were used for the modulation input. Thereby, we could tune $n_{\text{chips}}$ between 7 and 2047. The sequences were stored on a field-programmable-gate-array (FPGA) (AMD Artix-7 in Zynq 7000 XC7Z045 SoC on ZC706 Evaluation Board) and transmitted as differential signal to a dedicated EOM driver (iXblue Photonics (exail): DR-DG-20-HO) by an onboard serial \si{\giga\hertz}-transceiver of the ZC706 evaluation board. The FPGA could be programmed to transmit the m-sequences at different frequencies, for this work we mainly used $f_{\text{PRN}}\!=\!\SI{5}{\giga\hertz}$ and $f_{\text{PRN}}\!=\!\SI{10}{\giga\hertz}$.

For in-situ observation of the PRN sequence, the PRN modulated light was interferred with unmodulated light. Using a high-speed photo-diode (Thorlabs: RXM42AF, 42 GHz Photoreceiver) allowed to resolve the modulation. The EOM driving voltage could be optimized by observing the strength of the injected scattered light note in a live-spectrum.

\subsection{Michelson interferometer setup}
The Michelson interferometer setup had equal arm-lengths of around \SI{82.2}{\centi\meter}, one arm could be controlled in length with a PZT delay line. 
The other arm contained a PZT mirror for adjusting the phase, locking the interferometer and simulating (gravitational wave) signals. Light was split of the interferometer in the other arm by a low power reflectivity ($R_{\text{sc}}\!\approx\!\SI{20}{\percent}$) mirror to create a nominal and a parasitic beam path. The resulting maximum phase error in the Michelson was about \SI{0.0625}{\radian} and a power imbalance between the interferometer beams of around \SI{64}{\percent} was introduced. 
The parasitic beam was phase modulated by another PZT mirror and coupled back via the same low-reflectivity mirror. The delay $\tau_{sc}$ relative to the light in the interferometer could be adjusted with a linear translation stage.

The interferometer was locked and read out at mid-fringe by taking the difference between the light power measured at the anti-symmetric and symmetric port. 
The control-loop for this lock had a unity gain frequency of around \SI{5.5}{\kilo\hertz}, resulting in the interferometer being effectively free running at frequencies above \SI{100}{\kilo\hertz}. 
We injected two different sine-signals, one simulating a gravitational wave at $f_{\text{gw}} = \SI{172.4}{\kilo\hertz}$ with the piezo-actuator in the north-arm directly into the interferometer and one at $f_{\text{sc}} = \SI{170}{\kilo\hertz}$ to modulate the phase of the parasitic beam that simulates scattered light 
coupled into the east-arm.  As the scatter dynamics, limited by the strength of the PZT resonance at $f_{\text{sc}}$, did not move through a full fringe, the actual phase error $\varphi_{\text{err}}$ depended on the relative phase $\varphi_{\text{sc},0}$ of the scattered light. With the power $P_0$ in the interferometer and scattered light power $P_{\text{sc}}$ it was given by
\begin{equation}
    \varphi_{\text{err}}(t) = \sqrt{\frac{P_{\text{sc}}}{P_0}}\sin(\varphi_{\text{sc},0}+\delta\varphi_{\text{sc}}(t)).
\end{equation}
The dynamic was introduced by the injected modulation $\delta\varphi_{\text{sc}}(t)$ at $f_{\text{sc}}$. To achieve maximum phase error in each measurement even with $\varphi_{\text{sc},0}$ slowly fluctuating, a ramp was added to the PZT. This ensured that each measurement contained the maximal coupling.

\subsubsection{Recording and treatment of data}
Data was recorded by taking time series either with or without the PRN modulation active. 
These had four million data points sampled over two seconds with a sampling rate of \SI{2}{\mega\hertz}. 
From the recorded data, several spectra were computed using Welch's method with a Blackman-Haris window and \SI{50}{\percent} overlap. The reached suppression was calculated by comparing the peaks at $f_{\text{sc}}=\SI{170}{\kilo\hertz}$ between spectra calculated over the full time series measured with and without PRN-modulation. The upper and lower limits were calculated by comparing minimum and maximum values of the same peak in the spectra, this time calculated for 30 parts of the same length distributed evenly over the whole recorded time series.

\subsection{Power-recycled Michelson setup}

For the power-recycling a mirror was placed in the input path about \SI{10}{\centi\meter} before the central beam splitter to form a power-recycling cavity of \SI{0.922}{\meter} length or \SI{1.844}{\meter} roundtrip length. The interferometer was locked to an operating point resulting in destructive interference at the asymmetric port using a dither locking technique. The dither frequency was chosen to be \SI{172.4}{\kilo\hertz} to use a strong resonance of the piezo controlling the interferometers operation point. 
The laser was locked to the power recycling cavity by actuating on its frequency and using the PDH technique to lock the frequency to the cavity resonance. 
For the needed rf phase modulation of the incoming light field, a second fiber-coupled EOM (iXblue Photonics: NIR-MPX-LN-0.1) with a bandwidth of \SI{150}{\mega\hertz} was introduced. This EOM was driven by the a second dedicated EOM driver (iXblue Photonics (exail): DR-VE-0.1-MO).

To avoid signal overlap between scattered light signal and dither signal, the scattered light signal was injected at a frequency of \SI{135}{\kilo\hertz}, making use of a different piezo resonance.

\subsection{Linear cavity setup}
The experimental setup to demonstrate compatibility of tunable coherence and optical resonators consisted of a folded, linear cavity. This cavity was microscopically tuned in length with a PZT mirror, the folding mirror. It was set up using two mirrors with \SI{99.5}{\percent} power reflectivity as input and end mirrors, the input mirror was flat, the end mirror had a radius of curvature of \SI{5}{\meter}. Two photodetectors in transmission and reflection measured the transmitted and reflected laser power, respectively. The reflected signal was used to lock the cavity onto microscopic resonance using the PDH-technique, making use of the same EOM for rf-sideband modulation as described before. 
The Finesse of the cavity was around 696.

The initial round-trip length of the cavity, $l_{\text{cav}}$, was chosen to be $\SI{4.496}{\meter}$ in order to match the recoherence length of a 15~chips long sequence at $\SI{10}{\giga\hertz}$ PRN-modulation frequency ten times. To fine adjust the matching between both length, the PRN-frequency was tuned by changing the reference clock frequency for the transceiver on the FPGA which is simply proportional. 
Additionally, several other combinations of integer multiples of a PRN sequence fitting the cavity could be realized by using other PRN frequencies and sequences.

\subsubsection{Recording and treatment of data}
For the scanned measurement of the PRN-modulation resonance, the transmitted power and the current frequency of the FPGA-reference clock were recorded synchronously. From these recordings, the transmitted power in relation to PRN detuning could be recovered.

%

\end{document}